\documentclass[amsmath,amssymb,superscriptaddress,nobalancelastpage,prb,twocolumn,10pt]{revtex4-1}

\AtBeginDocument{%
    \newwrite\bibnotes
    \def\bibnotesext{Notes.bib}
    \immediate\openout\bibnotes=\jobname\bibnotesext
    \immediate\write\bibnotes{@CONTROL{REVTEX41Control}}
    \immediate\write\bibnotes{@CONTROL{%
    apsrev41Control,author="08",editor="1",pages="1",title="0",year="1"}}
     \if@filesw
     \immediate\write\@auxout{\string\citation{apsrev41Control}}%
    \fi
}%
\usepackage{svg}
\usepackage{amsmath}

\usepackage{graphicx}
\usepackage{varioref}
\usepackage{xr-hyper}
\usepackage{xcolor}
\usepackage{nicefrac}
\usepackage{xfrac}
\usepackage{hyperref}
\hypersetup{colorlinks,linkcolor=blue,urlcolor=blue,citecolor=orange}
\usepackage{ulem}
\usepackage{siunitx}
\usepackage{graphicx}
\usepackage{dcolumn}
\usepackage{bm}
\usepackage{braket}
\usepackage{wasysym}
\usepackage{textcomp}

\newcommand{\NTO}{Ni$_3$TeO$_6$}

\newcommand{\MTO}[1]{\texorpdfstring{#1$_3$TeO$_6$}{{#1}3TeO6}}

\newcommand{\JParameters}{$J_1$ = -9.27, $J_2$ = -12.90, $J_3$ = 1.22, $J_4$ = 0.13, $D_3$ = 0.65 and $D_{ani}$ = -0.27 meV}

\let\OLDthebibliography\thebibliography
\renewcommand\thebibliography[1]{
  \OLDthebibliography{#1}
  \setlength{\parskip}{0pt}
  \setlength{\itemsep}{0pt plus 0.3ex}
}

\begin{document}

\title{
Microscopic origin of field-induced multiferroicity in \texorpdfstring{Ni$_3$TeO$_6$}{Ni3TeO6}}

\author{J. Lass}
\affiliation{PSI Center for Neutron and Muon Sciences, 5232 Villigen PSI, Switzerland}
\affiliation{Laboratory for Quantum Magnetism, Ecole Polytechnique Fédérale de Lausanne (EPFL), CH-1015 Lausanne, Switzerland}
\affiliation{Nanoscience Center, Niels Bohr Institute, University of Copenhagen, DK-2100 Copenhagen \O , Denmark}

\author{R. Toft-Petersen}
\affiliation{Department of Physics, Technical University of Denmark, DK-2800 Kongens Lyngby, Denmark}

\author{A. A. Turrini} 
\affiliation{Laboratory for Quantum Magnetism, Ecole Polytechnique Fédérale de Lausanne (EPFL), CH-1015 Lausanne, Switzerland}
\author{\O. S. Fjellv\aa g} 
\affiliation{PSI Center for Neutron and Muon Sciences, 5232 Villigen PSI, Switzerland}
\affiliation{Department for Hydrogen Technology, Institute for Energy Technology, NO-2027 Kjeller, Norway}

\author{B. Decrausaz}
\affiliation{PSI Center for Neutron and Muon Sciences, 5232 Villigen PSI, Switzerland}
\affiliation{Physik-Institut, Universit\"{a}t Z\"{u}rich, Winterthurerstrasse 190, CH-8057 Z\"{u}rich, Switzerland}

\author{K. M. L. Krighaar}
\affiliation{Nanoscience Center, Niels Bohr Institute, University of Copenhagen, DK-2100 Copenhagen \O , Denmark}
\affiliation{PSI Center for Neutron and Muon Sciences, 5232 Villigen PSI, Switzerland}

\author{R. Sibille} 
\affiliation{PSI Center for Neutron and Muon Sciences, 5232 Villigen PSI, Switzerland}

\author{H. Jacobsen} 
\affiliation{PSI Center for Neutron and Muon Sciences, 5232 Villigen PSI, Switzerland}
\affiliation{European Spallation Source ERIC, Data Management and Scientific Computing, Asmussens Allé 305, 2800 Kongens Lyngby, Denmark}
\affiliation{Nanoscience Center, Niels Bohr Institute, University of Copenhagen, DK-2100 Copenhagen \O , Denmark}

\author{M. E. Kamminga} 
\affiliation{Nanoscience Center, Niels Bohr Institute, University of Copenhagen, DK-2100 Copenhagen \O , Denmark}
\affiliation{Condensed Matter and Interfaces, Debye Institute for Nanomaterials Science, Utrecht University, 3508 TA Utrecht, The Netherlands}

\author{C. R. Y. Andersen} 
\affiliation{National Centre for Nano Fabrication and Characterization, Technical University of Denmark, Building 307, Kgs. Lyngby 2800, Denmark}
\affiliation{Nanoscience Center, Niels Bohr Institute, University of Copenhagen, DK-2100 Copenhagen \O , Denmark}

\author{D. Sheptyakov} %
\affiliation{PSI Center for Neutron and Muon Sciences, 5232 Villigen PSI, Switzerland}

\author{J. Ollivier} 
\affiliation{Institut Laue-Langevin, 71 avenue des Martyrs, CS 20156, 38042 Cedex 9 Grenoble, France}

\author{U. B. Hansen} 
\affiliation{Institut Laue-Langevin, 71 avenue des Martyrs, CS 20156, 38042 Cedex 9 Grenoble, France}

\author{M. Enderle} 
\affiliation{Institut Laue-Langevin, 71 avenue des Martyrs, CS 20156, 38042 Cedex 9 Grenoble, France}

\author{A. M. L\"auchli} 
\affiliation{
PSI Center for Scientific Computing, Theory and Data, 5232 Villigen PSI, Switzerland}
\affiliation{Laboratory of Theoretical and Computational Physics, Ecole Polytechnique Fédérale de Lausanne (EPFL), CH-1015 Lausanne, Switzerland}

\author{H. M. Rønnow}
\affiliation{Laboratory for Quantum Magnetism, Ecole Polytechnique Fédérale de Lausanne (EPFL), CH-1015 Lausanne, Switzerland}

\author{K. Lefmann} 
\affiliation{Nanoscience Center, Niels Bohr Institute, University of Copenhagen, DK-2100 Copenhagen \O , Denmark}

\author{D. G. Mazzone} 
\affiliation{PSI Center for Neutron and Muon Sciences, 5232 Villigen PSI, Switzerland}

\date{\today}

\begin{abstract}
The continued demand for smaller, faster, and more energy-efficient technologies has driven condensed-matter research towards materials combining multiple responses within a single phase. Particularly interesting examples are multiferroic materials featuring cross-couplings between order parameters. \NTO\ is a prominent member of this material class, exhibiting a hysteresis-free transition under magnetic field which simultaneously affects the magnetic structure, electric polarization, and phonon frequencies. Here, we uncover the microscopic origin of this multiferroic transition using a combination of neutron scattering techniques. We establish the presence of a finite Dzyaloshinskii-Moriya exchange term giving rise to non-reciprocal, chiral spin waves which soften in magnetic field. Their condensation alters the magnetic structure which controls the electric polarization and drives an ionic displacement, thereby coupling the magnetic, electric, and lattice degrees of freedom. Our study demonstrates that neutron scattering is a powerful technique to elucidate microscopic interactions in multiferroic materials, informing the design of next-generation electronic devices.

\end{abstract}

\maketitle
Collective order in solid state materials refers to a symmetry-broken ground state in which a vast number of charges, spins, orbitals, or ions act in a coordinated manner. Examples include crystalline, antiferromagnetic (AFM), and ferroelectric order, yielding emergent phonon, magnon, and ferron quasiparticles, which are associated with the fluctuations of the respective finite order parameters. Multiferroic materials exhibit several collective orders in a single phase, and are of particular importance if the various order parameters are coupled.\cite{Eerenstein06,Vopson2015,Dong2019} In these cases the field-conjugate response of one degree of freedom affects several order parameters present in the phase, exemplified by a magnetic field also influencing the electric polarization and conversely an electric field altering the magnetization of a material. These properties are relevant for several technological applications, including modern electronics\cite{Nan2017,Pradhan2020,Spaldin2019,Ba2021}, bio-medical applications\cite{Gu2019,Smith2023}, and the field of photovoltaics and photocatalysis,~\cite{Spaldin2020,Chatterjee2022} which has led to a rapid increase in fundamental and applied research of multiferroicity in recent years. 

One family of materials exhibiting strong couplings between magnetic and charge degrees of freedom is the transition metal tellurates \MTO{M}, M = $\{$Co, Cu, Mn, Ni$\}$.\cite{Hudl2011, Ivanov2012, Wang2013, Singh2014, Singh2016, Herak2005, Norman2018, Bao2018, Shahee2023, Kisicek2024, Ivanov2011, Zhao2015, Ivanov2017, Zivkovic2010, Wu2010, Sankar2013, Oh2014, Kim2015, Yokosuk2015, Yokosuk2016, Skiadopoulou2017, Ghosh2018, Lass2020Ni3TeO6,Sondarva2021,Numan2022_2} Magnetic order is frequently stabilized in non-collinear magnetic structures enabling magnetoelectric effects. For instance, the cubic \MTO{Cu} hosts a cycloidal spin structure and features a magnetoelectric transition in applied magnetic field,~\cite{Herak2005, Norman2018, Bao2018, Shahee2023, Kisicek2024} and a helical magnetic order with fluctuating moments in the rhombohedral \MTO{Mn} yields a magnetoelectric coupling.~\cite{Zhao2015, Ivanov2017, Sondarva2021} Rhombohedral \NTO\ features a hysteresis-free transition at low temperature and $\mu_0H_c$ = 8.6 T applied along the $c$-axi (in the hexagonal unit cell description) across which the material reveals a multiferroic coupling between magnetic, charge, and lattice degrees of freedom, as evidenced by concurrent changes in the electric permittivity,~\cite{Oh2014,Yokosuk2016} lattice constants,~\cite{Kim2015,Numan2022_2} ion displacements,~\cite{Numan2022_2} phonon frequencies,~\cite{Yokosuk2015} and magnetic structure. ~\cite{Zivkovic2010,Oh2014,Lass2020Ni3TeO6}

In vanishing magnetic field \NTO~orders below a Néel temperature of $T_N =$ 52 K in a collinear AFM structure with wavevector $\mathbf{Q}_{AFM} = (0, 0, 1.5)$ in reciprocal lattice units (r.l.u.).~\cite{Ghosh2018,Sankar2013,Zivkovic2010,Skiadopoulou2017,Lass2020Ni3TeO6} The magnetic excitation spectrum consists of low-energy spin waves with minima in their dispersion relation at incommensurate (IC) positions $\mathbf{Q}^\pm_{IC} = (0, 0, 1.5\pm\delta)$ with $\delta\simeq0.18$, which has previously been interpreted using a $J_1$-$J_2$ model with frustrated next nearest-neighbor interactions.~\cite{Lass2020Ni3TeO6} The spin-wave gap of $\hbar\omega$ $\simeq1.2$ meV determined by inelastic neutron scattering is consistent with the lowest-energy electro-magnon observed by Raman scattering.~\cite{Skiadopoulou2017} Above the field-induced transition, the exact magnetic structure has remained elusive, but experimental results have established a helical magnetic order with wavevector $\mathbf{Q}_{IC}$ and magnetic moments rotating within the hexagonal $ab$-plane.~\cite{Lass2020Ni3TeO6}

Here, we report a comprehensive neutron scattering study in magnetic fields up to $\mu_0H$ = 9 T applied along the crystallographic $c$-axis to determine the nature of the field-induced multiferroic transition of \NTO. Using polarized neutron spectroscopy, we reveal the chiral nature of the spin waves through an asymmetric exchange interaction, and neutron diffraction measurements provide direct evidence for exchange striction. Based on these results we develop an effective magnetic Hamiltonian accounting for the low-energy dynamics of the material. The model accurately predicts the spin-wave dispersion, intensity modulation, and chirality of the lowest-energy spin-wave branches, alongside their magnetic field dependencies, whose condensation triggers the magnetic phase at the critical field. We conclude that the multiferroic coupling arises from a Dzyaloshinskii-Moriya (DM) interaction, providing a microscopic understanding of the field-induced multiferroicity in \NTO. The study also demonstrates that neutron scattering is a powerful tool to elucidate microscopic couplings in multiferroic materials.

\section{Results}
\begin{widetext}
    \begin{figure*}[!ht]
        \centering
        \includesvg[width=0.97\linewidth]{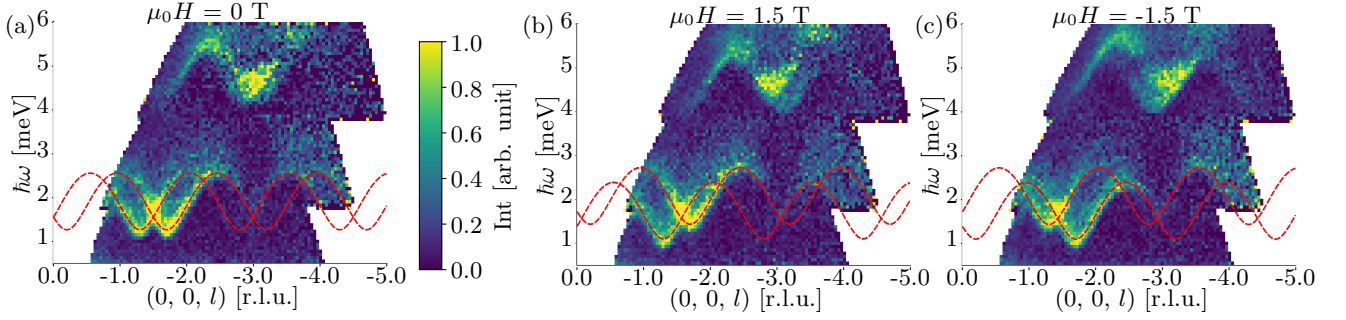}
        \caption{\textbf{Magnetic-field dependence of excitations along (0, 0, $\boldsymbol{l}$) in the collinear phase.} Inelastic neutron scattering results of \NTO\ at $T$ = 5 K plotted as function of energy transfer, $\hbar\omega$, and momentum transfer, $\mathbf{Q}$, along (0, 0, $l$) for (a) $\mu_0H$ = 0, (b) 1.5 T, and (c) -1.5 T applied along the $c$-axis using the hexagonal unit cell description. An integration width of 0.05 1/\AA, $\mathbf{Q}$-bin size of 0.02 1/\AA, and energy bin size of 75 $\mu$eV was used. The color-scale denotes the neutron scattering intensity. The spectra are over-plotted with the spin-wave model detailed in Sec.~\ref{sec:Model}. We remind the reader of the symmetry conversions (0, 0, -1.5 $\pm\delta$) = (0, 0, -3) + $\mathbf{Q}^\pm_{IC}$ and (0, 0, -3 $\pm\delta_2$) = (0, 0, -3) + $\mathbf{Q}^\pm_{IC2}$.}
        \label{fig:figure2}
    \end{figure*}
\end{widetext}

We start by presenting our neutron diffraction results followed by inelastic neutron scattering studies under magnetic field, and at zero field using a polarized neutron beam. We refer to Sec.~\ref{sec:methods} for experimental details.

\subsection{Crystal structure at zero magnetic field }

High-resolution neutron powder and single-crystal diffraction on the diffractometers HRPT\cite{Fischer2000} and Zebra at the Swiss Spallation Neutron Source SINQ, Paul Scherrer Institut (PSI), Switzerland confirm the reported nuclear structure of \NTO.~\cite{Newnham1967,Becker2006} The compound crystallizes in $R$3 (space group 146), which describes a noncentrosymmetric, chiral crystal structure. The structure is retained at least down to 1.5 K $<$ $T_N = 52$ K with lattice constants $a$ = 5.10183(1) and $c$ = 13.75231(6) \AA\ at $T$ = 1.5 K. The fractional ion positions are listed in Tab.~\ref{tab:HRPT_powder} with Ni$_1$ being fixed to restrict a sliding origin (the unit cell is shown in App. \ref{app:HRPT}). In accordance with previous reports, we find that the material orders below $T_N$ in a collinear AFM structure along the $c$-axis~\cite{Zivkovic2010} with $\mathbf{Q}_{AFM}$ = (0, 0, 1.5)~\cite{Lass2023} and ordered moment $\mu$ = 1.93(2)$\mu_\mathrm{B}$. No quantitative differences were observed in magnetic refinements, in which the magnetic moments of the different Ni sites were assumed to be independent or were constraint to the same amplitude.

\begin{table}[ht!]
    \centering
    \begin{tabular}{c|c|c|c|c}
    Ion & x & y & z & $\mu$ [$\mu_\mathrm{B}$] \\ \hline
      Ni$_1$ & 0 & 0 & 0 & 1.93(2)$^*$    \\
      Ni$_2$ & 0 & 0 & 0.2020(3) & 1.93(2)$^*$   \\
      Ni$_3$ & 0 & 0 & 0.4928(3) & 1.93(2)$^*$   \\
      Te & 0 & 0 & 0.6940(2)  & -- \\
      O$_1$ & 0.3283(4) & 0.0265(3) & 0.09812(2) & --   \\
      O$_2$ & 0.3303(5) & 0.3709(3) & 0.2703(3)  & --  \\
    \end{tabular}
    \caption{\textbf{Nuclear structure and magnetic moment sizes.} The fractional ion positions in the hexagonal unit cell setting and the magnetic moment sizes were determined at $T$ = 1.5 K using high-resolution neutron powder diffraction, and were confirmed by single crystal refinements (see App. \ref{app:HRPT} for details). The refinements were carried out with FullProf\cite{Fullprof} and Mag2Pol\cite{Qureshi2019}. $^*$The magnetic moments were constrained to equal amplitudes.}
    \label{tab:HRPT_powder}
\end{table}

In contrast to earlier studies,~\cite{Numan2022_2} we observe only tiny ion movements across $T_N$. Notably, we find that the oxygen positions remain unchanged, but determine a minor ($\delta z =$ 4$\cdot10^{-4}$) translation of the Ni$_2$ and Ni$_3$ ions relative to Ni$_1$ with a slightly larger absolute change for Te. The shifts correspond to 0.005 \AA~ and generate an electric polarization of -0.26(18) $\mu$C/cm$^2$ that is consistent with literature. \cite{Oh2014} (see App.~\ref{app:HRPT} for details). The largest angular change along the Ni-O-Ni superexchange bonds is 0.13(10)$^\circ$.

\subsection{Non-reciprocal, chiral spin waves in the collinear AFM phase}

\begin{figure}[!ht]
    \centering
    \includesvg[width=0.97\linewidth]{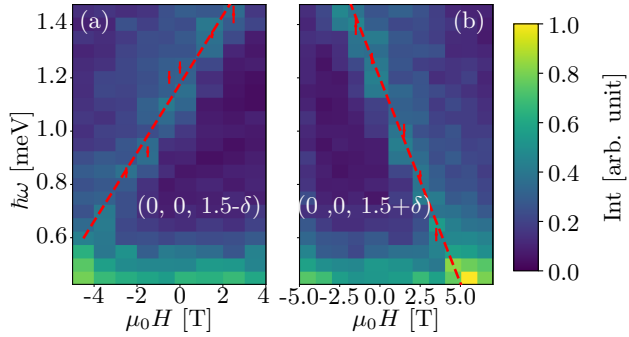}
    \caption{\textbf{Magnetic-field dependence of the magnon gap.} Magnon gap as function of applied magnetic field $\mu_0H$ $\parallel$ $c$ for (a) $\mathbf{Q}^-_{IC}$ = (0, 0, 1.5-$\delta$) and (b) $\mathbf{Q}^+_{IC}$ = (0, 0, 1.5+$\delta$) with $\delta=0.18$. We observe zero-field spin-wave gaps at $\hbar\omega$ = 1.19(2) meV with linear magnetic field dependencies of $\pm$0.144(8) meV/T (see dashed red lines).}
    \label{fig:TASPFieldDependence}
\end{figure}
The low-energy, magnetic excitation spectrum at $T$ = 5 K and zero field was measured on the multiplexing neutron spectrometer CAMEA\cite{Lass2023} at SINQ, PSI, and is shown in Fig.~\ref{fig:figure2}(a). The spectrum contains two sets of non-reciprocal magnon branches with minima at $\mathbf{Q}^\pm_{IC}$ and $\hbar\omega$ $\simeq$ 1.2 meV, and $\mathbf{Q}^\pm_{IC2}$ = (0, 0, $\pm\delta_2)$ with $\delta_2<0.05$ at $\hbar\omega$ $\simeq$ 4.5 meV. The magnetic excitations reveal a magnon bandwidth of $\simeq1.5$ meV along (0, 0, $l$), and extend to $\hbar\omega$ $\simeq9$ and 13 meV along ($h$, 0, 0) (not shown).
Under magnetic field $\mu_0H$ $\parallel$ $c$ the four magnon branches shift in energy, so that for positive fields the modes at $\mathbf{Q}^+_{IC}$ and $\mathbf{Q}^-_{IC2}$ soften while the ones at $\mathbf{Q}^-_{IC}$ and $\mathbf{Q}^+_{IC2}$ harden (see Fig.~\ref{fig:figure2}(b)). We observe that this behavior is reversed in negative field $\mu_0H$ $\parallel$ -$c$, meaning that the modes at $\mathbf{Q}^+_{IC}$ and $\mathbf{Q}^-_{IC2}$ harden while $\mathbf{Q}^-_{IC}$ and $\mathbf{Q}^+_{IC2}$ soften (see Fig.~\ref{fig:figure2}(c)).

The detailed magnetic-field dependence of the two low-energy spin-wave gaps was measured on the triple-axis spectrometer TASP\cite{Boni1996} at SINQ, PSI. The results shown in Fig.~\ref{fig:TASPFieldDependence} reveal equal-size zero-field gaps of $\hbar\omega$ = 1.19(2) meV and a linear field dependence with proportionality factor $\pm$0.144(8) meV/T. The Zeeman splitting is consistent with a magnetic moment of $\mu$ = 1.99(10)$\mu_\mathrm{B}$ for Ni$^{2+}$ (Landé g-factor $g_L$ = 5/4), which is in line with the magnetic moment determined through neutron diffraction. The magnetic field dependence of the spin-wave gaps suggests a complete magnon softening at $\mu_0H$ = $\pm$8.3(1) T for $\mathbf{Q}^\pm_{IC}$, respectively, which is close to the experimentally determined value of the multiferroic transition at $\mu_0H_c\simeq8.6$ T.~\cite{Lass2020Ni3TeO6,Yokosuk2015,Yokosuk2016,Oh2014,Kim2015}

\begin{widetext}
\begin{figure*}[!ht]
    \centering 
    \includesvg[width=0.97\linewidth]{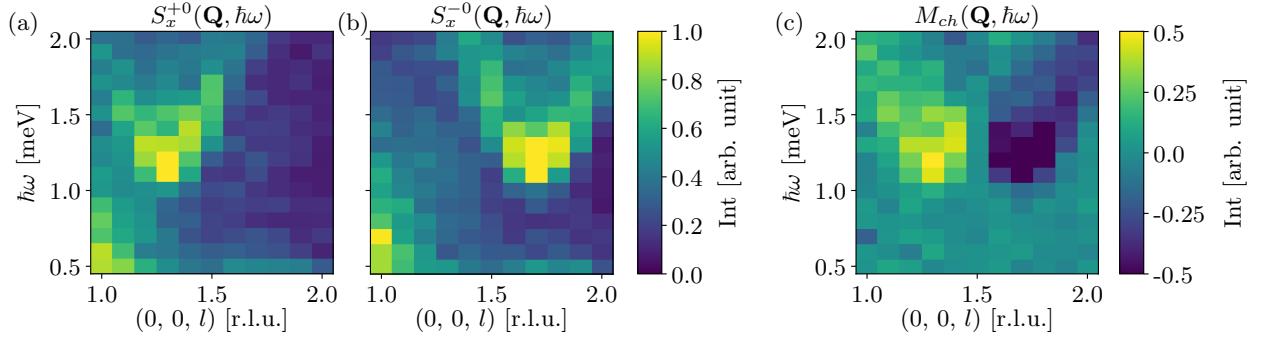}
    \caption{\textbf{Determination of spin-wave chirality.} Neutron scattering intensity of the two low-energy spin-wave branches along (0, 0, $l$) as function of energy transfer, $\hbar\omega$, measured at $T$ = 5 K using an instrument configuration with incident beam polarization. In panel (a) the neutron intensity associated with the cross section      S$_x^{+0}$($\mathbf{Q}$, $\hbar\omega$) is shown, which denotes an incident beam polarization along the scattering vector. (b) The corresponding intensity for     S$_x^{-0}$($\mathbf{Q}$, $\hbar\omega$), with incident beam polarization opposite to the scattering vector. (c) The chiral contribution to the magnetic neutron cross section determined by M$_{ch}(\mathbf{Q}, \hbar\omega) \propto \frac12\left(S_x^{+0}(\mathbf{Q}, \hbar\omega)-S_x^{-0}(\mathbf{Q}, \hbar\omega)\right)$ (see App.~\ref{app:Polarized} for details).
    }
    \label{fig:polarised}
\end{figure*}
\end{widetext}

The distinct sensitivity of the non-reciprocal magnon branches to the magnetic field direction suggests that they carry different chirality. We tested this hypothesis through polarized neutron scattering experiments using TASP\cite{Boni1996} at SINQ, PSI, and the triple-axis spectrometer IN20\cite{Kulda2002,IN20} at the Institut Laue-Langevin (ILL), Grenoble, France. On TASP we concentrated on the two low-energy magnons with minima at $\mathbf{Q}^\pm_{IC}$ and determined the chiral contribution, $M_{ch}$($\mathbf{Q}$, $\hbar\omega$), to the magnetic neutron cross section through a setup in which the incident beam was polarized. The setup probes the cross sections 
S$_x^{+0}$($\mathbf{Q}$, $\hbar\omega$) and S$_x^{-0}$($\mathbf{Q}$, $\hbar\omega$), so that M$_{ch}(\mathbf{Q}, \hbar\omega) \propto \frac12\left(S_x^{+0}(\mathbf{Q}, \hbar\omega)-S_x^{-0}(\mathbf{Q}, \hbar\omega)\right)$. We define S$_x^{if}(\mathbf{Q}, \hbar\omega)$ as the cross-section with initial spin state $i$ and final state $f$ with polarization along the scattering vector $\mathbf{Q}$. 0 signifies that the neutron spin is not analyzed (see App.~\ref{app:Polarized} for details). Our results are shown in Fig.~\ref{fig:polarised}, revealing that the two spin-wave branches carry opposite chirality, $i.e.$, the branch with minimum $\mathbf{Q}^-_{IC}$ has positive chirality in contrast to $\mathbf{Q}^+_{IC}$ which is negative. In a separate experiment on IN20 we determined $M_{ch}$($\mathbf{Q}$, $\hbar\omega$) around $\mathbf{Q}^\pm_{IC}$ and $\mathbf{Q}^\pm_{IC2}$ with $\hbar\omega$ = 1.5 and 5 meV, respectively, using a setup in which both incident and final neutron polarization were determined. This experiment confirms the results shown in Fig.~\ref{fig:polarised} and determines a positive (negative) chirality of the spin-wave branch with minimum at $\mathbf{Q}^-_{IC2}$ ($\mathbf{Q}^+_{IC2}$) (see App.~\ref{app:Polarized} for details).

\subsection{Excitation spectrum in the helical phase}
\begin{figure}
    \centering
    \includesvg[width=0.97\linewidth]{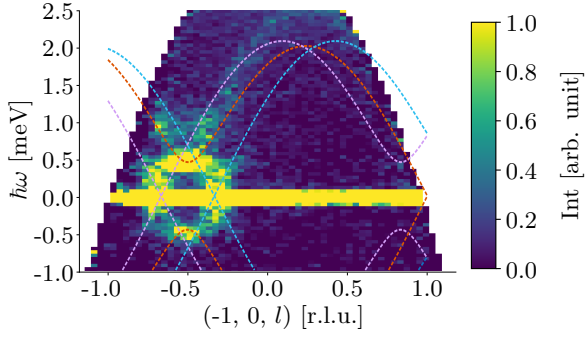}
    \caption{\textbf{Excitation spectrum along (-1, 0, $\boldsymbol{l}$) in the helical phase.} Inelastic neutron scattering results of \NTO\ at $T$ = 5 K and $\mu_0H$ $\parallel$ $c$ at $\mu_0H$ = 9 T, revealing two Goldstone modes emerging from $\mathbf{Q}^\pm_{IC}$ and a gapped spin wave with minimum at $\mathbf{Q}^\pm_{AFM}$. An integration width of 0.015 1/\AA\ along $h$ and 0.03 1/\AA\ along $k$ with bin sizes of 0.03 1/\AA\ along $l$, energy bin size of 20 $\mu$eV was used. The color-scale denotes the neutron intensity. The spectrum is overplotted with the linear spin-wave predictions using the model described in Sec.~\ref{sec:Model}.}
    \label{fig:IN5}
\end{figure}

A magnetic field applied along the $c$-axis of \NTO\ triggers a magnetic phase transition at $\mu_0H_c\simeq8.6$ T. The low-energy excitation spectrum in the helical phase was measured at $\mu_0H$ = 9 T using the time-of-flight neutron spectrometer IN5\cite{Ollivier2004} at ILL. In Figure~\ref{fig:IN5} we show the corresponding neutron scattering spectrum, revealing two Goldstone modes emerging from the incommensurate positions $\mathbf{Q}$ = (-1, 0, -0.5$\pm\delta$) with $\delta\simeq 0.18$, which are symmetry-related to $\mathbf{Q}^\pm_{IC}$. We also observe a gapped mode ($\hbar\omega$ $\simeq$ 0.45 meV) with minimum at $\mathbf{Q}$ = (-1, 0, -0.5), which is symmetry-related to $\mathbf{Q}_{AFM}$ = (0, 0, 1.5), the magnetic wavevector at $|H|$ $<$ $H_c$. We find that the gapped magnon branch extends to $\hbar\omega\simeq$ 2.3 meV along (-1, 0, $l$) (see Fig.~\ref{fig:IN5}) and $\hbar\omega\simeq$ 11.5 meV along ($h$, 0, -0.5) (not shown). In contrast, the intensity of the Goldstone modes reduces rapidly with increasing energy transfer along (-1, 0, $l$) (see Fig.~\ref{fig:IN5}), but they remain observable up to $\hbar\omega\simeq$ 7.5 meV along ($h$, 0, -0.5$\pm\delta$) (not shown).

\section{Discussion} 
We will first discuss our experimental results with respect to the magnetic degrees of freedom. This will provide a phenomenological picture of the field-induced transition at $\mu_0H_c$ and an explanation for the hysteresis-free nature of this transition. We will also elaborate on the implications on the magnetic structure at $\mu_0H$ $>$ $\mu_0H_c$. We will then present our effective microscopic model, which will elucidate the multiferroic coupling between charge, spin, and lattice degrees of freedom. Lastly, we will discuss the limitations of the model. 

\subsection{Field-induced condensation of chiral spin waves}

Collective magnetic order in \NTO\ manifests below $T_N$ = 52 K ($\mu_0H$ = 0 T) in a collinear AFM structure with wavevector $\mathbf{Q}_{AFM}$ = (0, 0, 1.5) .~\cite{Ghosh2018,Sankar2013,Zivkovic2010,Skiadopoulou2017,Lass2020Ni3TeO6} Our neutron scattering results show that the magnetic excitation spectrum consists of non-reciprocal, chiral spin waves with low-energy branches possessing minimal spin gaps at $\mathbf{Q}^\pm_{IC}$ = (0, 0, 1.5 $\pm$ 0.18) (see Fig.~\ref{fig:figure2}(a)). We observe that one of the spin-wave branches ($\mathbf{Q}^+_{IC}$ for $\mu_0H$ $\parallel$ $c$) softens under magnetic field and condenses into the elastic line at $\mu_0H$ = 8.6 T. As a result of detailed balance $S(-\mathbf{Q},-\hbar\omega) = S(\mathbf{Q},\hbar\omega)e^{\frac{-\hbar\omega}{k_BT}}$, the other spin-wave branch ($\mathbf{Q}^-_{IC}$ for $\mu_0H$ $\parallel$ $c$) softens in the energy gain channel of the magnetic excitation spectrum ($\hbar\omega<$ 0). Together they establish magnetic order at $\mathbf{Q}^\pm_{IC}$ independent of the magnetic field direction. Thus, we conclude that the field-induced transition is driven by the condensation of these low-energy magnon branches, so that below $\mu_0H_c$ they act as dynamic precursors of the magnetic order parameter above $\mu_0H_c$. We refer the reader to Ref. \onlinecite{Gitgeatpong2017} for an example in which the energy-gain and energy loss-spectrum of non-reciprocal spin waves have been reported.

The condensation of the chiral magnons in magnetic field provides an intuitive picture for the helical structure at $\mu_0H$ $>$ $\mu_0H_c$. Here, the magnetic excitation spectrum consists of two Goldstone modes emerging from $\mathbf{Q}^\pm_{IC}$ (see Fig.~\ref{fig:IN5}), which may be associated with fluctuations of the helical phase angle defining the relative, in-plane rotation of the canted magnetic moments~\cite{Lass2020Ni3TeO6}. Polarized inelastic neutron scattering can test this hypothesis in principle, but the required high magnetic field prevents such experiments with existing neutron instrumentation. The gapped mode centered at $\mathbf{Q}_{AFM}$ acts as a dynamic precursor of the magnetic order parameter below $\mu_0H_c$ = 8.6 T, similar to the non-reciprocal spin waves in the collinear AFM state. We hypothesize that the gapped mode at $\mathbf{Q}_{AFM}$ softens upon decreasing field $\mu_0H$ $>$ $\mu_0H_c$, which can be verified in future inelastic neutron scattering experiments.

The condensation of low-energy magnons at $\mu_0H_c$ epitomizes a cross over between the collinear and helical magnetic long-range orders, rationalizing the hysteresis-free nature of the transition. We also note that finite instrument resolution effects yield a field range close to $\mu_0H_c$ over which the two magnetic long-range orders at $\mathbf{Q}_{AFM}$ and $\mathbf{Q}^\pm_{IC}$ have been reported to coexist\cite{Lass2020Ni3TeO6} (see App~\ref{app:Coexistence} for details).

\subsection{Effective microscopic Hamiltonian}\label{sec:Model}

Having established that the magnetic field-induced transition in \NTO\ is driven by the condensation of the lowest-energy spin-wave branches, allows us to develop an effective model consistent with the experimental observations of these quasiparticles. We consider a magnetic Hamiltonian of the form
\begin{equation}\label{eq:model}
    \mathcal{H} = \sum_{ij} \mathbf{S}_i J_{ij,D} \mathbf{S}_j + \sum_i \left(D_\mathrm{ani} S_{i,z}^2 +\mu\mu_\mathrm{0}\mathbf{H} \mathbf{S}_i\right),
\end{equation}
where $\mathbf{S}_i$ is the spin operator on site $i$, $J_{ij}$ is the interaction matrix between sites $i$ and $j$, $D_\mathrm{ani}$ denotes the single ion anisotropy along the $c$-axis. The first sum uniquely assigns the nearest-neighbor bonds, the second sum runs over all ions. The interaction matrix is described by
\begin{equation}
    J_{ij,D} = \begin{pmatrix}
        J_{ij} & D_{z,ij} & -D_{y,ij} \\ 
        -D_{z,ij} & J_{ij} & D_{x,ij} \\ 
        D_{y,ij} & -D_{x,ij} & J_{ij}
    \end{pmatrix},
\end{equation}

where $\mathbf{D}_{ij}$ = ($D_{x,ij}$, $D_{y,ij}$, $D_{z,ij}$) defines the DM vector, which can be separated into $\mathbf{D}_{ij} = D_{ij}\hat{\mathbf{\epsilon}}_{ij}$. $\hat{\epsilon}_{ij} = \hat{\mathbf{r}}_i\times\hat{\mathbf{r}}_j$ is the unit vector calculated through the two vectors $\hat{\mathbf{r}}_i$ and $\hat{\mathbf{r}}_j$, connecting Ni-ions on sites $i$ and $j$ with the neighboring ligand oxygen ion to define the Ni-O-Ni superexchange pathway.

\begin{table*}[!ht] 
    \centering
    \begin{tabular}{c|c|c|c|c}
         & Ions & Ni distance & $\angle$Ni-O-Ni & $\hat{\epsilon}_{ij}$   \\
        &      &     [Å]     &         [deg] & [1/\AA]    \\ \hline  
$J_1$ & Ni$_2$-Ni$_3$ & 3.0037 & 88.1 & (0.567, 0.161, 0.807) \\
$J_2$ & Ni$_1$-Ni$_2$ & 2.7885 & 98.4 & -- \\ 
$J_3$ & Ni$_1$-Ni$_2$ & 3.4497 & 117.3 & (0.468, -0.229, 0.853) \\
$J_4$ & Ni$_1$-Ni$_3$ & 3.6767 & 125.3 & (0.300, -0.571, 0.764) \\
    \end{tabular}
    \caption{\textbf{Considered interactions of the effective magnetic Hamiltonian.} The table shows the considered interactions, the ions involved in the interaction, the interaction distance, the superexchange bond angles, and possible $\hat{\epsilon}_{ij}$ based on our neutron diffraction results at $T$ = 5 K. The interaction pathways are shown in App. \ref{app:HRPT}.}
    \label{table:interactions}
\end{table*}

In Table~\ref{table:interactions} we list the four nearest-neighbor interactions of our model, together with their interaction distances, superexchange bond angles, and $\hat{\epsilon}_{ij}$, obtained from our powder diffraction experiments at $T$ = 5 K. These results suggest an interaction hierarchy with dominating ferromagnetic $J_2$ $<$ 0 interaction, followed by $J_1$ $<$ 0 and AFM $J_3$ $>$ $J_4$ $>$ 0. We expect a finite easy-axis anisotropy $D_{ani}$ $<$ 0 to account for the zero-field spin-wave gap in the collinear AFM structure. Our preliminary simulations showed that a finite splitting of the spin-wave branches along (0, 0, $l$) occurs only with a DM interaction on AFM bonds. Thus, we considered a finite DM term $D_3$ on $J_3$, which is enabled by an off-centered O$^{-2}$ ion with respect to $\mathbf{e}_{3,2}$ connecting the Ni$^{+2}$ ions at the Ni$_2$ and Ni$_3$ positions (see App. \ref{app:HRPT}). 

The six parameters $J_1$-$J_4$, $D_3$, and $D_{ani}$ were refined against 1066 fitted magnon positions stemming from reciprocal lattice positions in the ($h$, 0, $l$) and ($h$, $k$, 0) planes and at magnetic fields $\mu_0H$ = 0, $\pm$1.5, 4, and 7 T in the collinear AFM phase. Using the linear spin-wave theory package SpinW,~\cite{Toth2015} we obtained 
\JParameters.

In Figure~\ref{fig:figure2} we overplot the dispersion relation obtained from our effective model with the experimental data measured on CAMEA. The model accurately predicts the spin-wave gap, bandwidth, and non-reciprocal splitting of the low-energy magnon branches. It also models their intensity modulation and chirality (see App.~\ref{app:Model}). Under magnetic field the model further accounts for the magnetic field dependence of the two magnon branches, correctly predicting the experimental spectra at $\mu_0H$ = $\pm$ 1.5, 4.5 and 7 T. We attempted to predict the magnetic excitation spectrum in the incommensurate phase at 9 T $>$ $\mu_0H_c$. The simulations were carried out in an extended cell of 1 x 1 x 12 unit cells to approximate the incommensurate wavevector $\mathbf{Q}^\pm_{IC}$ and the spin structure was obtained through a temperature annealing process (see App.~\ref{app:Model} for details). We find that the data measured on IN5 are satisfactorily predicted (see Fig.~\ref{fig:IN5}), but only if a magnetic field of $\mu_0H$ = 9.75 T is assumed. We attribute this discrepancy to an insufficient model of the magnetic structure which is currently unattainable, an unsatisfactory approximation of $\mathbf{Q}^\pm_{IC}$ through the cell extension, and an incorrect assumption that the magnetic interactions are retained across the field-induced transition (see also Sec. \ref{sec:MultiferroicCoupling}). 

Despite the minor inconsistency, the effective model provides an appropriate description of the lowest-energy magnon branches in \NTO. However, we mention that the effective model is limited to these magnon branches and does not accurately predict the spin-wave branches at higher-energy transfers centered at $\mathbf{Q}^\pm_{IC2}$ ($\mu_0H<\mu_0H_c$). One reason is the short-coming to satisfactorily account for differences in the local environment around the three Ni-sites, which leads to variations in the magnetic anisotropy. Deviations can also arise from further-neighbor and higher-order interactions that are neglected in our minimal model. Both limitations may be reflected in an overestimated $D_3$ term, which potentially absorbs some of these effects. 

\subsection{Multiferroic coupling}\label{sec:MultiferroicCoupling}

Our model allows for an effective description of the low-energy spin-wave branches in \NTO\ and their behavior in magnetic field. Notably, the model establishes a finite DM term $D_3$, which splits the two otherwise degenerate branches into non-reciprocal, chiral spin waves. The finite DM interaction establishes the previously missing link between the magnetic, charge, and lattice degrees of freedom in \NTO. Experimental evidence for such a coupling is found through a variety of techniques, including our neutron scattering results, measurements of the dielectric constant,~\cite{Oh2014,Yokosuk2016} infrared,~\cite{Yokosuk2015} and Raman spectroscopy.~\cite{Skiadopoulou2017} Notably, our neutron diffraction results determine tiny Ni-ion shifts across $T_N$ below which we detect the DM interaction. This is in agreement with phonon anomalies\cite{Yokosuk2015} suggesting a change in the superexchange interaction. The movement of the Ni-ions along the $c$-axis also induces an electric polarization of -0.26(18) $\mu$C/cm$^2$ that is consistent with dielectric measurements which have reported a pyroelectric effect across $T_N$.~\cite{Oh2014} Raman measurements reveal the presence of electro-magnons at $\hbar\omega$ = 1.30, 1.57 and 4.37 meV,~\cite{Skiadopoulou2017} which match the values of the spin-wave gaps and their crossing point (see Fig.~\ref{fig:figure2}). These results connect the charge and lattice degrees of freedom with the magnetic DM interaction.

At $\mu_0H_c$ \NTO\ can be controlled by a magnetic or an auxiliary electric field along the $c$-axis.~\cite{Oh2014} In the former case, the low-energy, non-reciprocal magnons condense into the elastic line if the magnetic field is increased across $\mu_0H_c$. The condensation prompts the collinear AFM spin alignment to change into a helical order.~\cite{Lass2020Ni3TeO6} The altered magnetic structure yields an electric polarization $P_{ij}$ $\propto$ $\mathbf{e}_{i,j}$ x ($\mathbf{S}_{i}$ x $\mathbf{S}_{j}$) between sites $i$ and $j$,~\cite{Tokura2014} which is absent for collinear AFM spin structures but is finite in helical spin orders. The change in electric polarization prompts tiny ion movements, whose existence has been confirmed indirectly through phonon anomalies occurring across the field-induced transition.~\cite{Yokosuk2015} The accompanied change in the DM interaction arising from adjusted Ni-O positions also provides an explanation for the discrepancies between the experimental results and our theoretical model at $\mu_0H$ = 9 T. 

Conversely, the application of an electric field controls the electric polarization, which affects the lattice degrees of freedom through electrostriction and influences the DM term. The resulting modification of $D_3$ can further soften the low-energy magnons near $\mu_0H_c$, promoting its condensation and enabling an effective electric-field control of the magnetic degrees of freedom. This mechanism is supported by our spin-wave model, which show that an increase of the $D_3$ term leads to an appreciably softer low-energy magnon below $\mu_0H_c$. We note that the same argument applies in reverse, when starting from the helical phase above $\mu_0H_c$.

\section{Conclusion}
\NTO\ displays one of the largest reported magneto-electric couplings across a field-induced transition at $\mu_0H_c$ = 8.6 T.~\cite{Oh2014} Using a combination of comprehensive neutron scattering experiments and theoretical modeling, we determined the microscopic origin of this multiferroic coupling. At zero field we observe non-reciprocal, chiral spin waves below $T_N$, which arise from a finite Dzyaloshinskii-Moriya term connecting the magnetic, charge, and lattice degrees of freedom. Under magnetic field the lowest-energy magnons soften and condense at $\mu_0H_c$, which alters the magnetic structure. The transition from a collinear AFM to a helical structure changes the electric polarization which is accompanied by tiny ionic-movements. Thus, a magnetic field alters the magnetic structure and affects the electric polarization in \NTO. Conversely, an electric field controls the electric polarization, which modifies the Dzyaloshinskii-Moriya term through electrostriction. At fields close to $\mu_0H_c$ this effect leads to the condensation of the lowest magnon branch promoting changes in the magnetic structure, so that an electric field can be used to control electric and magnetic properties of the material.

The interplay between spin, charge, and lattice degrees of freedom established by the Dzyaloshinskii-Moriya interaction, in principle enables the control of the magnetic and electric properties through elastic effects. We hypothesize that strain applied along the $c$-axis at magnetic fields close to $\mu_0H_c$ alters the Dzyaloshinskii-Moriya interaction, also affecting the field-induced transition in \NTO. Its critical field is governed by a combination of the zero-field spin gap, the Zeeman term, and the Dzyaloshinskii-Moriya interaction strength. Partial substitution of the magnetic ion affects the Zeeman term, which would also impact the magnetic anisotropy that governs the size of the magnon gap. The Dzyaloshinskii-Moriya interaction strength is determined by local Ni-O-Ni superexchange bonds, so that innovative material engineering may find opportunities to control these mechanisms and reduce the critical field of the material. Although the multiferroic transition in \NTO\ will remain unattainable for direct technological applications, our study attests that neutron scattering is a powerful technique to elucidate microscopic interactions in multiferroic materials which is a prerequisite for technological breakthroughs. 

\section{Methods}\label{sec:methods}

\subsection{Material preparation and characterization.}
The neutron powder diffraction experiment was carried out using a double-wall Vanadium can with inner diameter of 4 mm containing $m$ = 8 g of \NTO\ powder. The single crystal neutron diffraction experiment was performed on a $m$ $\simeq$ 10 mg high-quality single crystal, which was mounted on an aluminum pin with GE-vanish. All inelastic neutron scattering experiments were performed on the same $m$ = 306 mg sample. The crystalline quality was confirmed using X-ray and neutron Laue diffraction. The crystal features a clear facet orthogonal to the $c$-axis and extends 4 x 4 x 3 mm$^3$ along the $a$-, $b$-, and $c$-axis, respectively. The crystal was mounted with GE-vanish on a 2 mm thick Al plate for experiments in which the $c$-axis resided in the horizontal scattering plane, and was mounted on a thin Al pin when the $c$-axis was vertical.

\subsection{Neutron powder and single crystal diffraction}

Neutron powder diffraction was performed on the thermal powder diffractometer HRPT\cite{Fischer2000} at SINQ, PSI, using an ILL-type Orange cryostat. Neutron diffraction patterns were measured at different temperatures for about 3 hours per temperature using the High Intensity mode with neutron wavelength $\lambda$ = 1.1545 \AA. A representative data set is shown in Fig.~\ref{fig:HRPT}. At base temperature ($T$ = 1.5 K), the Medium Resolution mode with $\lambda$ = 2.4489 \AA~was used. We employed a scattering angle coverage from 6$^\circ$ to 160$^\circ$ for all setups. Using the FullProf suite\cite{Fullprof}, the data were refined against the nuclear structure of \NTO, an approximated cubical impurity NiO phase, including its magnetic contribution, the structure of the Vanadium can, and when applicable against the magnetic structure of \NTO.

Single crystal neutron diffraction was conducted at the thermal neutron diffractometer Zebra at SINQ, PSI using a neutron wavelength $\lambda$ = 1.383 \AA\ obtained with a Ge(220) monochromator. Motorized slits were adjusted on a few Bragg peaks, about 30 cm before the sample and in front of a $^3$He tube used as a point detector. An open HUBER cradle combined with a dedicated $T$ = 5 - 300 K cryostat was used. Data were collected using omega scans and integrated intensities determined by fits to a Gaussian function and a linear background, followed by Lorentz correction. Data were refined using Mag2Pol\cite{Qureshi2019}, and results of the nuclear and magnetic refinements are shown in Fig.~\ref{fig:ZebraDiffraction}.

\subsection{Single crystal neutron spectroscopy} 

The dynamic processes in \NTO\ were studied on the following four neutron spectrometers: the triple-axis spectrometer TASP\cite{Boni1996} at SINQ, PSI, the multiplexing spectrometer CAMEA\cite{Lass2023} at SINQ, PSI, the triple-axis spectrometer IN20\cite{Kulda2002} at ILL, and time-of-flight spectrometer IN5\cite{Ollivier2004} at ILL. The sample was aligned either with ($h$, $h$, $l$), ($h$, 0, $l$), or ($h$, $k$, 0) in the horizontal scattering plane and was measured at a temperature of $T = $ 5 K. 

\subsubsection{CAMEA: Experiments in horizontal and vertical fields}
Inelastic neutron scattering experiments on CAMEA were performed in multiple configurations. We employed the horizontal-field magnet MA02 to measure the excitation spectrum of \NTO\ for $\mu_0H$ = -1.5, 0, and 1.5 T for incoming energies $E_i$ = 5.5, 7.1, 8.7, 9.3, and 9.6 meV, with scattering angles $2\theta =$ -42 and -38$^\circ$ for $E_i < 9.3$ meV and $2\theta = $ -34 and 38$^\circ$ for $E_i \ge 9.3$ meV. Sample scans were performed over 80$^\circ$ in steps of 1$^\circ$, centered around (0, 0, 1.5) counting between 1 and 3 minutes/point. The resulting data shown in Fig.~\ref{fig:figure2} have been treated with MJOLNIR\cite{Lass2020MJOLNIR} software (software version 1.3.3, doi: 10.5281/zenodo.13772982). We re-normalized the data attenuated by a structural pillar of the horizontal-field magnet, so that incoherent elastic scattering remained constant across sample rotation scans. The magnet was setup such that the pillar only crossed the direct beam. 

The excitation spectrum of \NTO\ was also studied in the horizontal ($h$, $k$, 0) scattering plane for fields $\mu_0H$ $<$ 7 T using the vertical MB11 magnet. Sample rotation scans in the hexagonal plane were performed in 1$^\circ$ steps for a range of 80$^\circ$ covering (1, 0, 0) and (0, 1, 0).

In both experiments we observed sharp spin waves, validating the assumption that all sample-rotation-independent contributions to the excitation spectra stem from parasitic scattering. We collapsed these background features over momentum transfer, $\|\textbf{Q}\|$, and energy bins of 0.03 1/\AA~and 0.1 meV, respectively, and subtracted them from the data, similar to what has been done in Ref. \onlinecite{Facheris2022}. 

\subsubsection{TASP: Incident beam-polarization setup}
We performed an inelastic neutron scattering experiment with incident beam polarization on TASP employing the MuPad setup reported in Refs. \onlinecite{Roessli2002a,Janoschek2007}. We used a final energy of $E_f = $ 9.6 meV to determine the chirality of the low-energy spin-wave branches (see App.~\ref{app:Polarized} for details). The neutron response was measured as a function of (0, 0, $l$) and energy transfer, $\hbar\omega$, in steps of 0.1 r.l.u. and 0.1 meV, respectively. Each ($\mathbf{Q}$, $\hbar\omega$)-point was measured with incident neutron polarization along (+) and opposite to (-) $\mathbf{Q}$ for a total of 2.5 min per point.

\subsubsection{TASP: Experiment in horizontal field}
The field dependence of the low-energy magnon gap was investigated with the horizontal field magnet MA7 using a crystal alignment in the ($h$, 0, $l$) plane. Energy scans with $E_f = $ 9.6 meV were performed at (0, 0, 1.5 $\pm$0 .02) over an energy range $\hbar\omega$ = 0.45 - 1.45 meV and field range $\mu_0H$ = -4.5 - 6 T in steps of 1 T. Every point was measured for around 5 min.

\subsubsection{IN20: Polarized neutron scattering experiment}
A polarized inelastic neutron scattering experiment was performed on the triple-axis spectrometer IN20,~\cite{Kulda2002} using a Heusler-Heusler setup together with a Helmholtz coil enabling rotation of the beam polarization. A velocity selector suppressed higher harmonic scattering and we used $k_f$ = 2.662 1/\AA\ with a PG filter. The setup was used to measure the chirality of the spin-wave branches around $\mathbf{Q}^\pm_{IC}$ and $\mathbf{Q}^\pm_{IC2}$ with $\hbar\omega$ = 1.5 and 5 meV (see App.~\ref{app:Polarized} for details). The neutron response was measured in steps of 0.05 and 0.1 r.l.u. along (0, 0, $l$) counting for 5 and 8 min/point at $\hbar\omega$ = 1.5 and 5 meV, respectively.

\subsubsection{IN5: Experiment at high magnetic field}
Low-energy excitation spectra on the time-of-flight spectrometer IN5 were measured at $\mu_0H$ = 4.5 and 9 T using neutron wavelengths $\lambda$ = 2 and 5 \AA. We employed sample rotation scans over 55$^\circ$ in steps of 1$^\circ$ with 10 min/point for $\lambda$ = 5 \AA. The scans with $\lambda$ = 2 \AA~were performed over 77$^\circ$ in 1$^\circ$ steps with 8 min/point. The data was treated using the HORACE\cite{Ewings2016a} software on the ILL VISA cluster and visualized through MJOLNIR.~\cite{Lass2020MJOLNIR} 

\section*{Acknowledgments}
We acknowledge Helmut Berger for synthesizing large \NTO\ single crystal appropriate for inelastic neutron scattering experiments and Maria Retuerto for the \NTO\ powder used for the neutron diffraction study. We thank O. Zaharko, for fruitful discussions. We thank the Paul Scherrer Institut and the Institut Laue-Langevin for the allocated beamtime on CAMEA (proposal numbers 20201264 and 20230918), TASP (proposal numbers 20210151 and 20212668), Zebra (proposal number 20202298), HRPT (proposal number 20230397), IN5 (proposal number 4-01-1784\cite{ILL-DATA.4-01-1784}), and IN20 (proposal number TEST-3527\cite{ILL-DATA.TEST-3527}).

\section*{Author contributions}

J.L. and K.L. initiated the project and planned it together with D.G.M. and A.A.T. The samples were synthesized by H.B., M. R. and C.R.Y.A. Experiments were performed by J.L., D.G.M., K.L., A.T.T., R.T.-P., K.M.L.K., B.D., D.S., H.J., J.O., U.B.H., M.E., C.R.Y.A., R.S., Ø.S.F., M.E.K. and H.M.R. The data were analyzed by J.L., D.G.M., B.D., R.S. and D.S. J.L. and D.G.M. interpreted the results together with A.M.L, K.L., H.M.R., Ø.S.F., and M.E.K. J.L. and D.G.M. wrote the manuscript with inputs from all co-authors.

\appendix

\section{High-resolution neutron powder and single crystal diffraction}\label{app:HRPT}
\begin{widetext}
    \begin{figure*}
    \includesvg[width=0.97\linewidth]{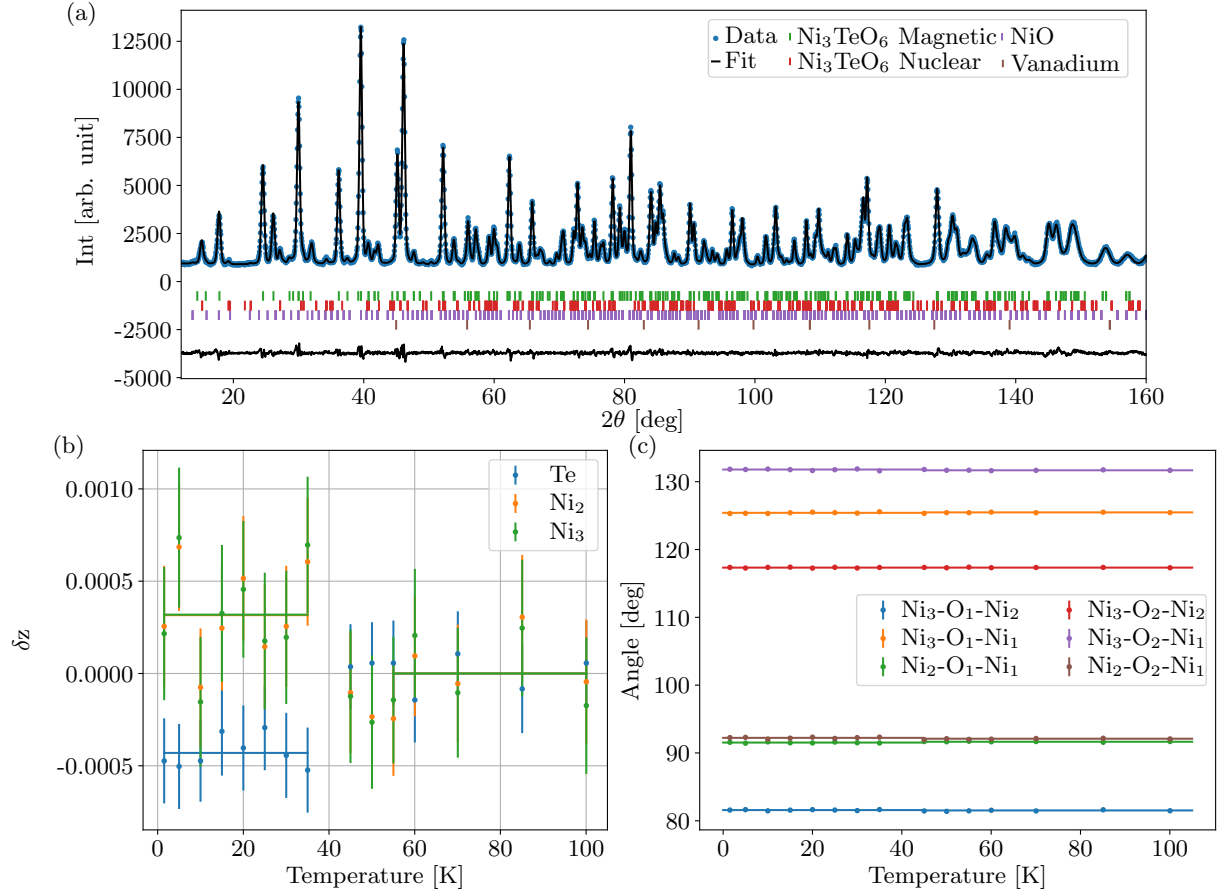}
        \caption{\textbf{Neutron powder diffraction and determined parameters.} (a) Diffraction pattern measured with $\lambda = 1.1545$ \AA\ at $T$ = 1.5 K, showing experimental data and refined model containing structural and magnetic \NTO\ phases, Vanadium can and a NiO impurity. Ticks below graph correspond to positions of the phases (b) Temperature dependence of fractional position along the $c$-axis with Ni$_1$ kept fixed to restrain a sliding origin. Lines signify weighted average above and below T$_N$ = 52~K. (c) Interaction path angles as function of temperature.}
        \label{fig:HRPT}
    \end{figure*}
\end{widetext}

The temperature dependence of lattice parameters and fractional positions were determined using neutron powder diffraction. The data measured at $\lambda = 1.1545$ \AA\ and $T$ = 1.5 K are shown in Fig.~\ref{fig:HRPT}(a) and were analyzed in a combined refinement of the \NTO\ nuclear structure (see Fig. \ref{fig:UnitCellAndJs}), the magnetic structure, the nuclear structure of the Vanadium can and a NiO impurity phase. The temperature dependence of the lattice constants and the oxygen positions were minimal, so they were neglected in the following analysis. In contrast, a significant change in the fractional positions of the nickel and tellurium ions were observed, amounting to 0.00032(17), 0.00030(17), and -0.00041(11) for Ni$_2$, Ni$_3$, and Te, respectively (see Fig.~\ref{fig:HRPT}(b) and Fig.~\ref{fig:HRPT}(c)). Notably, the interaction path angle of the Ni$_2$ - O$_2$ - Ni$_1$ superexchange changes by 0.13(10)$^\circ$. We specify that the ion movements were obtained through the average fractional positions for temperatures above $T =$ 55 K and were compared to those below $T =$ 45 K, so that the region around T$_N$ was omitted. The change in electric polarization of -0.26(18) $\mu$C/cm$^2$ was calculated using the electric charge of the nickel and tellurium ions, and the \NTO\ unit cell volume.

\begin{figure}[!ht]
    \centering
    \includesvg[width=0.97\linewidth]    {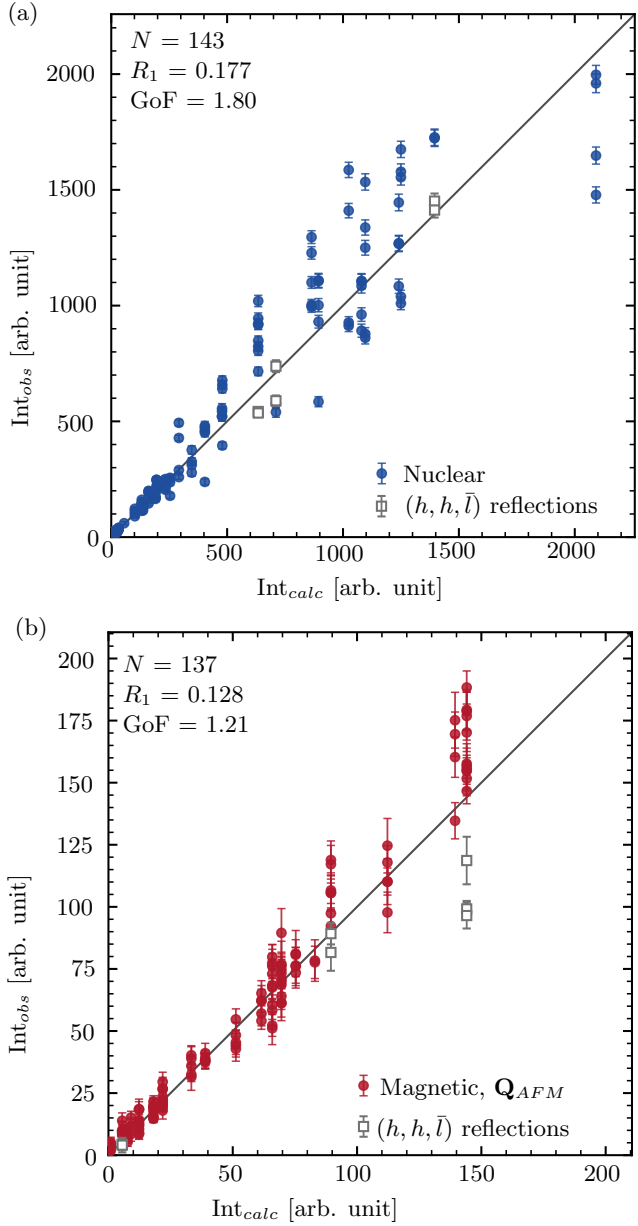}
    \caption{\textbf{Single-crystal neutron diffraction refinements.} Refinement of the single-crystal neutron diffraction results in the AFM phase at $T$ = 5 K ($\mu_0H$ = 0 T) performed on Zebra. (a) Comparison between observed and calculated integrated intensities for the nuclear structure and (b) magnetic structure. }
    \label{fig:ZebraDiffraction}
\end{figure}

\begin{widetext}
    \begin{figure*}
    \centering
\includesvg[width=0.57\linewidth]{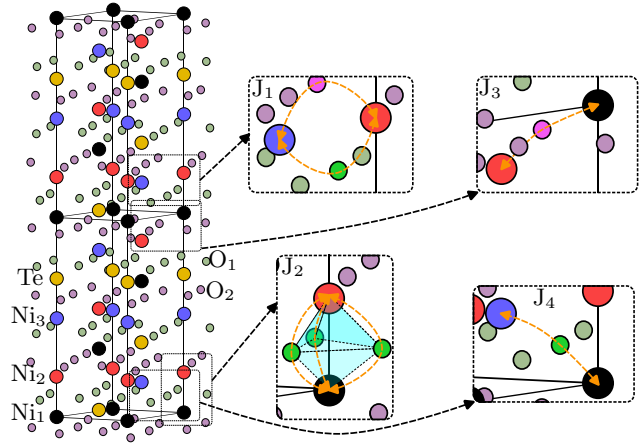}
        \caption{\textbf{Structural unit cell and interaction paths.} Nuclear structure showing Ni$_1$, Ni$_2$, Ni$_3$, Te, O$_1$, and O$_2$ as refined from powder and single crystal diffraction together with the interaction paths used in the effective model (see~\ref{sec:Model} for details). Interaction pathways are highlighted by the yellow arrows connecting two Ni ions through the brightly colored O ligands. }
        \label{fig:UnitCellAndJs}
    \end{figure*}
\end{widetext}

\section{Polarized neutron spectroscopy}\label{app:Polarized}
Following the standard convention,\cite{Boothroyd2020} we define the reference coordinate system so that $\hat{x}$ is along the incoming beam, $\hat{y}$ the orthogonal vector in the horizontal scattering plane, and $\hat{z}$ points vertically upwards in a right-handed coordinate system. Restricting ourselves to neutron polarization along $\mathbf{Q}$, and omitting the explicit dependence on $(\mathbf{Q},\hbar\omega)$, yields

\begin{align}
    S_x^{++}&\propto N+BG_{NSF}\\
    S_x^{--}&\propto N+BG_{NSF}\\
    S_x^{+-}&\propto M_{yy}^\perp+M_{zz}^\perp+M_{ch}+BG_{SF}\\
    S_x^{-+}&\propto M_{yy}^\perp+M_{zz}^\perp-M_{ch}+BG_{SF},
\end{align}
where S$_x^{if}$ represents the neutron initial, $i$, and final, $f$, spin state, $N$ the nuclear contribution, $BG_{SF}$ and $BG_{SF}$ the spin- and non-spin-flip background, respectively, $M_{yy}^\perp$ and $M_{zz}^\perp$ are the orthogonal components of the magnetic contribution along y and z, respectively, and $M_{ch}$ is the chiral component of the magnetic cross section. $M_{ch}$ can be obtained in a fully polarized neutron experiment such as the one performed on IN20 through
\begin{equation}
    M_{ch}\propto\frac12\left(S_x^{+-}-S_x^{-+}\right).
\end{equation}

In experiments with incident beam polarization only, such as on TASP, we obtain
\begin{align}
    S_x^{+0}&=S_x^{++}+S_x^{+-} \notag\\&\propto N+BG_{NSF}+ M_{yy}^\perp+M_{zz}^\perp+M_{ch}+BG_{SF} \\
    S_x^{-0}&=S_x^{-+}+S_x^{--} \notag\\&\propto N+BG_{NSF} +M_{yy}^\perp+M_{zz}^\perp-M_{ch}+BG_{SF},
\end{align}
so that
\begin{equation}
    M_{ch}\propto\frac12\left(S_x^{+0}-S_x^{-0}\right),
\end{equation}
similar to the fully polarized setup. 

\begin{figure*}[ht]
    \centering
    \includesvg[width=0.95\linewidth]{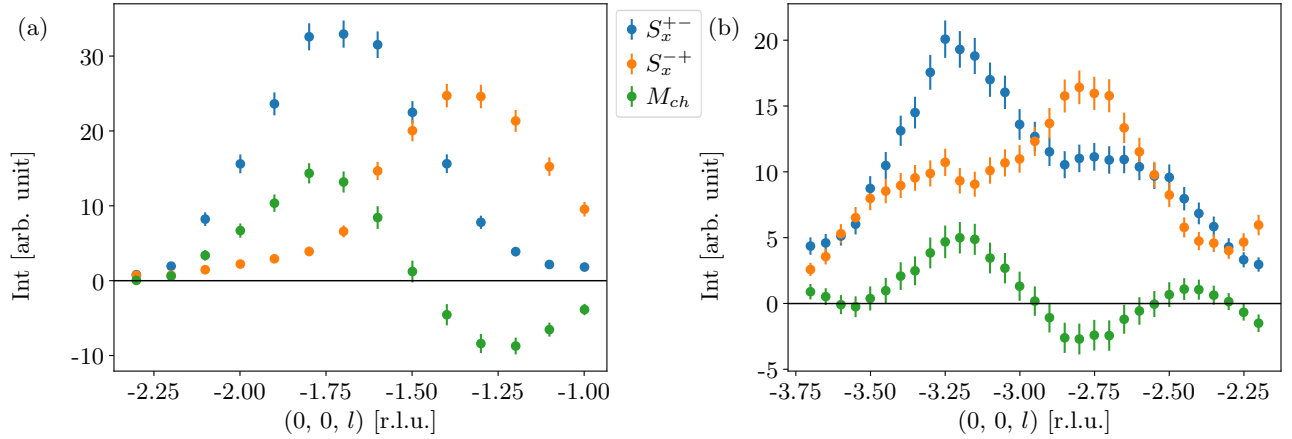}
    \caption{\textbf{Polarized neutron spectroscopy.} Inelastic scattering intensity at constant energy transfers of (a) $\hbar\omega$ = 1.5 meV and (b) $\hbar\omega$ = 5 meV from the polarized setup on IN20 showing neutron intensities associated to S$_x^{+-}$, S$_x^{-+}$, and the calculated intensity associated with chirality M$_{ch}$. We find that $\mathbf{Q}^+_{IC}$ = (0, 0, -1.5+$\delta$) and $\mathbf{Q}^+_{IC2}$ = (0, 0, -3+$\delta_2$) have negative chirality, and  $\mathbf{Q}^-_{IC}$ = (0, 0, -1.5-$\delta$) and $\mathbf{Q}^-_{IC2}$ = (0, 0, -3-$\delta_2$) positive chirality.}
    \label{fig:IN20Appendix}
\end{figure*}

The experimental results of the polarized experiment on IN20 are shown in Fig.~\ref{fig:IN20Appendix}.

\section{Origin of co-existence phase}\label{app:Coexistence}

A previous neutron scattering study reported a field range around $\mu_0H_c$, over which collinear AFM and helical order coexist (see Fig. 2 in Ref. \onlinecite{Lass2020Ni3TeO6}). Our inelastic neutron scattering results allow clarifying that this coexistence arises from finite instrument resolution effects. The referred experiment studied the magnetic Bragg peaks at $\mathbf{Q}_{AFM}$ and $\mathbf{Q}^\pm_{IC}$ using a triple-axis spectrometer with incident and final energy $E_i = E_f = $ 3.6 meV. The configuration yields an energy resolution of $\simeq$50 $\mu$eV around the elastic line, within which an unambiguous discrimination between static and dynamic processes is impossible. Based on the magnetic-field dependence of the lowest-energy magnon (see Fig.~\ref{fig:TASPFieldDependence}) we infer that the reported results did not allow for a clear assignment of the magnetic ground state over a field range of $\simeq$ 0.33 T, matching the field range over which the coexsitance was observed.

\section{Spin-wave simulations}\label{app:Model}

\begin{figure*}[!htb]
    \centering
    \includesvg[width=0.95\linewidth]{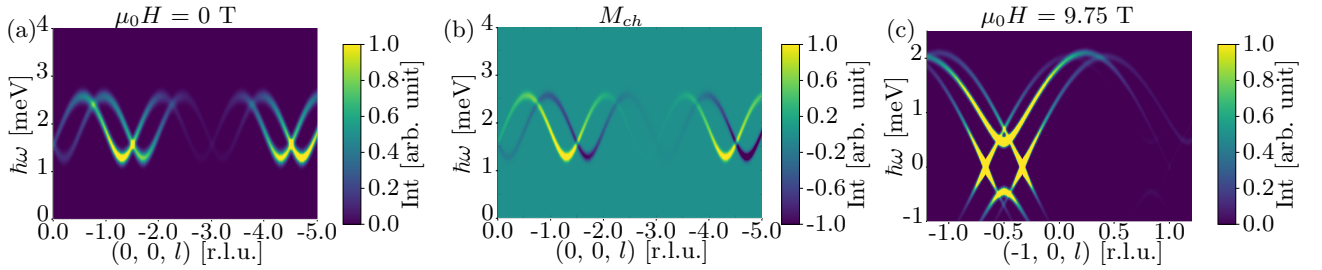}
    \caption{\textbf{Spin-wave simulations of effective model.} Predicted inelastic neutron scattering intensity at $\mu_0H$ = 0 T in (a) $M_{ch}$ in (b). (c) Excitation spectrum calculated for $\mu_0H$ = 9.75 T which approximates the experimental data acquired at $\mu_0H$ = 9 T.}
    \label{fig:TheoryPredictions}
\end{figure*} 

The experimental data were analyzed using the effective Hamiltonian defined in Eq.~\ref{eq:model} with finite $J_1$-$J_4$, $D_3$, and $D_{ani}$. The six parameters were refined against 1066 fitted magnon positions stemming from reciprocal lattice positions in the ($h$, 0, $l$) and ($h$, $k$, 0) planes and at magnetic fields $\mu_0H$ = 0, $\pm$1.5, 4, and 7 T. Using the linear spin-wave theory package SpinW,~\cite{Toth2015} we obtained 
\JParameters. The effective g-factor was kept isotropic at 2.0. Figures~\ref{fig:TheoryPredictions}(a) and~\ref{fig:TheoryPredictions}(b) show the predicted neutron scattering intensity at zero magnetic field and chiral contribution to the magnetic neutron cross section, $M_{ch}$, respectively (see Figs.~\ref{fig:figure2}(a) and~\ref{fig:polarised}(c) for comparison with experimental results). The excitation spectrum at $\mu_0H$ = 9.75 T has been calculated by extending the cell to 1 x 1 x 12 unit cells (see~\ref{fig:TheoryPredictions}(c) and Fig.~\ref{fig:IN5} for comparison to the experimental data). The magnetic structure was obtained through an annealing process for which we initiated a spin structure with substantial magnetic moment component along the $c$-axis in accordance with the determined low-field structure, and a small component randomly oriented in the hexagonal plane.

\newpage
\bibliography{Ni3TeO6}
\end{document}